\documentclass{IEEE-Con-Sys-mag}

\usepackage{amsmath}

\usepackage{caption}
\usepackage{subcaption}

\usepackage{pgf}
\usepackage{pgfplots}
\usepackage{tikz}
\usepgfplotslibrary{groupplots,fillbetween}
\usepackage{pgfplotstable}

\usepackage{url}
\usepackage{graphicx,xcolor}
\usepackage{verbatim}
\usepackage{soul}  \makeatletter
\newcommand{\verbatimfont}[1]{\def\verbatim@font{#1}}\makeatother
\verbatimfont{\ttfamily\small}
\usepackage[compress]{cite}

\newcommand{\bi}{\begin{itemize}}\newcommand{\ei}{\end{itemize}}
\newcommand{\be}{\begin{equation}}\newcommand{\ee}{\end{equation}}
\newcommand{\bee}{\begin{enumerate}}\newcommand{\eee}{\end{enumerate}}
\newcommand{\bea}{\begin{eqnarray}}\newcommand{\eea}{\end{eqnarray}}
\newcommand{\beas}{\begin{eqnarray*}}\newcommand{\eeas}{\end{eqnarray*}}
\newcommand{\bc}{\begin{center}}\newcommand{\ec}{\end{center}}

\definecolor{neurips}{RGB}{0,114,189}
\definecolor{icra}{RGB}{217,83,25}
\definecolor{cdc}{RGB}{119,172,48}

\makeatletter
\newcommand{\labelname}[1]{\def\@currentlabelname{#1}}\makeatother

\jvol{XX}
\jnum{XX}
\paper{XX}
\jmonth{August}
\jname{IEEE CONTROL SYSTEMS}
\pubyear{2023}

\begin{document}

\title{What is the Impact of Releasing Code with Publications?\stitle{Statistics from the Machine Learning, Robotics, \break and Control Communities}}

\author{SIQI ZHOU$^{1}$, LUKAS BRUNKE$^{1, 2}$, ALLEN TAO$^{2}$, ADAM W. HALL$^2$, FEDERICO PIZARRO BEJARANO$^2$, \break JACOPO PANERATI$^2$, AND ANGELA P. SCHOELLIG$^{1,2}$}
\affil{$^1$Technical University of Munich, $^2$University of Toronto}

\maketitle

\dois{}{}

\chapterinitial{O}pen-sourcing research publications is a key enabler for the reproducibility of studies and the collective scientific progress of a research community. As all fields of science develop more advanced algorithms, we become more dependent on complex computational toolboxes---sharing research ideas solely through equations and proofs is no longer sufficient to communicate scientific developments. Over the past years, several efforts have highlighted the importance and challenges of transparent and reproducible research; code sharing is one of the key necessities in such efforts. 
In this article, we study the impact of code release on scientific research and present statistics from three research communities: machine learning, robotics, and control. We found that, over a six-year period (2016-2021), the percentages of papers with code at major machine learning, robotics, and control conferences have at least doubled. Moreover, high-impact papers were generally supported by open-source codes. 
As an example, the top 1\% of most cited papers at the Conference on Neural Information Processing Systems (NeurIPS) consistently included open-source codes. 
In addition, our analysis shows that popular code repositories generally come with high paper citations, which further highlights the coupling between code sharing and the impact of scientific research.
While the trends are encouraging, we would like to continue to promote and increase our efforts toward transparent, reproducible research that accelerates innovation---releasing code with our papers is a clear first step.

\section{Introduction}
\label{sec:intro}

Reproducibility in research is critical to verifying the reliability of scientific ideas and thereby enabling scientific progress. 
Reproducibility, however, has become increasingly challenging in many disciplines. As an example, optimization is used in many different academic works, but the details of the optimizers are often not shared. Though the choices of libraries and hyper-parameters are usually not necessary to evaluate the novelty of a particular contribution, without these details, it can be extremely difficult to reproduce, verify, adopt, or compare against the original scientific work. Reproducible research is an indisputable cornerstone of innovation.

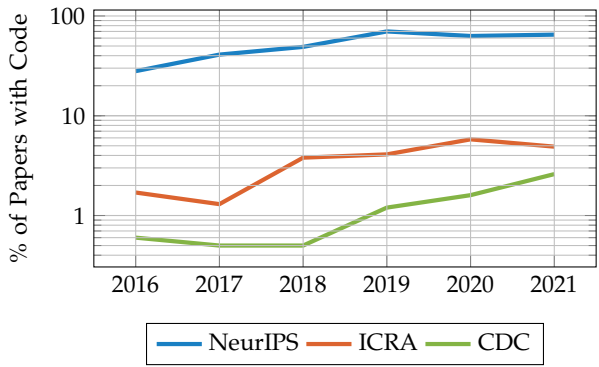
\begin{figure}[t]
    \centering
\begin{tikzpicture}
    \begin{groupplot}[
    group style = {group size = 1 by 1, horizontal sep=1.5cm, vertical sep=1.5cm,},
    width = 1.0\columnwidth, 
    height = 5cm,
    xmin=2016, 
    xmax=2021, 
ymode=log,
    grid=both, clip=false,
axis on top=true,
    clip marker paths=true,
    enlarge x limits=0.1,
legend style={at={(rel axis cs: 0.5,-0.3)}, anchor=center, legend columns = 3},
    x tick label style={
        /pgf/number format/.cd,
1000 sep={},
fixed,
        zerofill=true,
            precision=0,
            /tikz/.cd
        },
y tick label style={/pgf/number format/1000 sep=\,},
    log base 10 number format code/.code={$\pgfmathparse{10^(#1)}\pgfmathprintnumber{\pgfmathresult}$}]

    \nextgroupplot[
        xtick={2016, 2017, 2018, 2019, 2020, 2021},
        ylabel={\% of Papers with Code},
        ylabel near ticks,
]

            \addplot [
                color=neurips, 
                ultra thick,
                opacity=0.9,
            ] table [
                skip first n=1, 
                x index = {0}, 
                y index = {3}, 
                col sep=comma,
                each nth point=1,
            ]{./paperwithcode.csv};
            
            \addlegendentry{NeurIPS}
            
            \addplot [
                color=icra, 
                ultra thick,
                opacity=0.9,
            ] table [
                skip first n=1, 
                x index = {0}, 
                y index = {7}, 
                col sep=comma,
                each nth point=1,
            ]{./paperwithcode.csv};
            
             \addlegendentry{ICRA}
            
            \addplot [
                color=cdc, 
                ultra thick,
                opacity=0.9,
            ] table [
                skip first n=1, 
                x index = {0}, 
                y index = {11}, 
                col sep=comma,
                each nth point=1,
            ]{./paperwithcode.csv};
            
            \addlegendentry{CDC}

    \end{groupplot}
\end{tikzpicture}     \caption{Percentage of published papers with open-source code (with logarithmic y-axis) from a major conference in each of the three communities: (1) machine learning: The Conference on Neural Information Processing Systems (NeurIPS), (2) robotics: The IEEE International Conference on Robotics and Automation (ICRA), and (3) control: The IEEE Conference on Decision and Control (CDC). 
    }
    \label{fig:papers-with-code-summary}
\end{figure}

In recent years, numerous efforts have encouraged reproducibility and open-source code (OSC) in research. For instance, in the machine learning community, where large datasets and software infrastructures are already available, OSC is becoming common practice (see~\autoref{fig:papers-with-code-summary}). This trend was especially accelerated by recent reproducibility research efforts (e.g., reproducibility challenges and reproducibility checklists)~\cite{joelle2020}.

Papers with OSC facilitate scientific progress because they are easier to reproduce and benchmark against each other.
Coupled with high-quality datasets~\cite{newman2009data}, OSC can shape scientific fields (e.g., the MNIST, ImageNet, and CIFAR-10/CIFAR-100 datasets~\cite{mnist,imagenet,Krizhevsky2009LearningML} in machine learning). These datasets were instrumental to the deep learning revolution of the 2010s and have been collectively cited by over 30,000 research works~\cite{paperswithcode}.
In the robotics community, we have seen similar efforts, such as the Robot Operating System (ROS)~\cite{quigley2009ros}. ROS provides standardized interfaces for robot experiments, toolboxes~\cite{rtb}, and benchmark suites~
\cite{wang2019benchmarking}, encouraging fair comparisons and improving reproducibility. Moreover, in the control community, the subject of reproducible research has been raised in editorials and workshops~\cite{how2018control,how2015}.

Competitions often encourage OSC and reproducibility. We have also observed a growing number of competitions and challenges at major conferences. The NeurIPS conferences have included competition tracks since 2017; the number of competitions grew from 16 in 2020 to 23 in 2021.
Robot challenges have been part of ICRA for over a decade. In 2022, ICRA had a total of ten competitions, including the DodgeDrone~\cite{dodgedrone} and the ``Safe Robot Learning Competition''~\cite{iros-comp} (IROS 2022), that provided fully open-source codebases and encouraged the submission of public code solutions to promote open comparisons and reproducibility. 

These past efforts to encourage reproducible research highlight its importance to scientific progress and form the basis for trustworthy evaluations, comparisons, and further extensions of published results~\cite{the_turing_way2022}. We must continue to promote and increase our efforts in reproducible and transparent research to accelerate innovation; providing OSC with publications will be imperative to supporting fair evaluation and promoting further research.

In this article, we investigate the current status and the impact of OSC in scientific research. In particular, we present the OSC statistics from three research communities: \textit{(i)} machine learning, \textit{(ii)} robotics, and \textit{(iii)} control. 
The field of machine learning is leading in data-driven approaches and reproducibility discussions. 
Publications in this field often rely on a public dataset or simulation environment, and reproducing published results hinge on the availability of adequate computational resources. 
In contrast, the robotics community heavily relies on algorithmic designs and emphasizes hardware evaluations. Associated publications often require hardware-specific code that often cannot be easily transferred to other research groups' experimental hardware setups.
Finally, control is a traditionally theoretic field but is observed to have increased reliance on more complex and data-based techniques that require additional numerical or experimental demonstrations.
After presenting the statistics of OSC, we provide insights on the broader impact of code release and include lessons learned for encouraging further efforts to promote reproducible research.
We acknowledge that the differences in sharing OSC are partly due to the nature of the research conducted in the three communities, but the diversity also allows us to obtain transferrable insights in developing effective strategies that further encourage reproducibility collectively beyond the specialized domains.

 \section{Open-Source Code Statistics}

To keep the discussion concise, we selected one representative conference from machine learning, robotics, and control theory based on its $h5$-index.
The $h$-index estimates the impact of a conference's publication output and is given by the largest number, $h$, such that at least $h$ articles from that conference were cited at least $h$ times each~\cite{Hirsch2005}. The digit five in $h5$ indicates that this $h$-index only considers publications from the last five complete calendar years~\cite{googlescholarh5index}.
For the field of machine learning, we select the Conference on Neural Information Processing Systems (NeurIPS)~(Ranking in artificial intelligence: \#1 International Conference on Learning Representations~(ICLR)  with h5=286, \#2 NeurIPS with h5=278~\cite{googlescholarML}). 
Although ICLR has a slightly higher $h5$-index than NeurIPS, we have chosen to focus on NeurIPS to analyse the machine learning community to remain consistent with previous studies on reproducibility~\cite{joelle2020}.
For the field of robotics, we select the conference with the highest h5-index: the IEEE International Conference on Robotics and Automation (ICRA)~(Ranking in robotics: \#1 ICRA with h5=116, \#2 IEEE/RSJ International Conference on Intelligent Robots and Systems with h5=80~\cite{googlescholarRobotics}). 
For the field of control theory, we again select the conference with the highest h5-index: the IEEE Conference on Decision and Control~(CDC)~(Ranking in automation and control theory: \#1 CDC with h5=44, \#2 American Control Conference with h5=43~\cite{googlescholarControl}).

Measuring a publication's impact is multifaceted and challenging~\cite{Ravenscroft2017}. 
For the sake of simplicity, we focus on more readily available metrics that measure the academic impact. A publication's citation count is a frequently used quantity to assess academic impact~\cite{Ravenscroft2017}. 
For rigorousness, we use citation count data obtained exclusively from Semantic Scholar, as 
different sources for citation data often provide varying citation counts for the same publication.

The citation count is a cumulative quantity that is typically non-decreasing over time. 
Generally, publications that have been published later have had less time to collect citations.
Therefore, we always compare works published during the same year. 
However, this does not account for pre-prints released earlier on open-access platforms~(e.g., arXiv). 
We remark that proceeding publication dates typically differ among the selected conferences. Therefore, different citation counts for certain works can also be due to the varying release dates. 
Finally, the number of citations may also be impacted by the number of conference submissions or, more generally, the size of the specific research community since more conference submissions can yield publications with greater citation counts. 
We highlight that the numbers of submissions and publications at NeurIPS~(2021: 9122 submissions, 2344 accepted~\cite{neurips-num-submissions}), ICRA~(2021: 3877 submissions, 1690 accepted~\cite{icra-num-submissions2}), and CDC~(2021: 1735 submissions, 1097 accepted~\cite{cdc-2021-final-program}) differ.

All data presented in this article was collected from public sources. We have open-sourced the collected data and the associated analysis software that this article is based on~\cite{osc-repo}.
We refer the reader to the appendix for details on the collection process, see~\nameref{sec:appendix}.  \section{Status of Open-Source Code}
In this section, we give an overview of the current status of OSC at NeurIPS, ICRA, and CDC. The percentages of publications with OSC at the three conferences are summarized in~\autoref{fig:papers-with-code-summary}~(using a logarithmic y-axis). 
There exists an increasing trend of publications with OSC at NeurIPS, ICRA, and CDC.
However, the percentage of publications including OSC varies among the three conferences.
For the years from 2016 to 2021, \emph{publications at NeurIPS are ten and twenty times more likely to include OSC compared to publications at ICRA and CDC, respectively}.  

\begin{figure*}[!tb]
    \centering
    \begin{tikzpicture}
    \begin{groupplot}[
    group style = {group size = 3 by 1, horizontal sep=0.25cm, vertical sep=1.5cm,},
    width = 0.8\columnwidth, 
    height = 8cm,
    xmin=2016, 
    xmax=2021, 
ymin=0, 
ymax=2500,
    grid=none, clip=false,
axis on top=true,
    clip marker paths=true,
    legend pos=north west,
    x tick label style={
        rotate=60,
        /pgf/number format/.cd,
1000 sep={},
fixed,
        zerofill=true,
            precision=0,
            /tikz/.cd
        },
]

    \nextgroupplot[
        title={NeurIPS},
        xtick={2016, 2017, 2018, 2019, 2020, 2021},
ybar, enlarge x limits=0.25,
ylabel={Number of Publications},
        ylabel near ticks,
        ]

            \coordinate (A) at (axis cs:2016,1320);
            \coordinate (B) at (axis cs:2016,775);
            \coordinate (O1) at (rel axis cs:0,0);
            \coordinate (O2) at (rel axis cs:1,0);

            \addplot [
                bar shift=0pt,
                bar width=16pt,
                color=black!20, 
                fill=black!30,
                ultra thick,
                opacity=0.5,
] table [
                skip first n=1, 
                x index = {0}, 
                y index = {4}, 
                col sep=comma,
                each nth point=1,
            ]{./paperwithcode.csv};
            \addlegendentry{Total}

            \edef\cstmlst{
            "27.6\%",
            "40.5\%",
            "49.1\%",
            "69.7\%",
            "63.1\%",
            "65.5\%",
            }
            \addplot [
                bar shift=0pt,
                bar width=8pt,
                color=neurips,
                fill=neurips,
                ultra thick,
nodes near coords={\scriptsize \pgfmathsetmacro{\mystring}{{\cstmlst}[\coordindex]}\mystring},
            ] table [
                skip first n=1, 
                x index = {0}, 
                y index = {1}, 
                col sep=comma,
                each nth point=1,
            ]{./paperwithcode.csv};
            \addlegendentry{With Code}

\coordinate (C) at (axis cs:2018.5, 100);
            \coordinate (O3) at (rel axis cs:0,0);
            \coordinate (O4) at (rel axis cs:0,1);
            \draw [orange, ultra thick ,sharp plot,dashed,] (O3 -| C) -- (O4 -| C) node[right,pos=0.87,align=left] {\scriptsize NeurIPS\\ \scriptsize reprod. program};

            \draw [black!50,  thick ,sharp plot,dashed,] (A -| O1) -- (A -| O2) node[rotate=90,above,pos=0.99] {\scriptsize \hspace{4.5em} avg. 1320};
            \draw [neurips,  thick,sharp plot] (B -| O1) -- (B -| O2) node[rotate=90,above,pos=0.99] {\scriptsize \hspace{4.5em} avg. 775};

	 \nextgroupplot[
        title={ICRA},
        xtick={2016, 2017, 2018, 2019, 2020, 2021},
ybar, enlarge x limits=0.25,
ylabel near ticks,
        yticklabel=\empty, 
        ]

            \coordinate (A) at (axis cs:2016,964);
            \coordinate (B) at (axis cs:2016,38);
            \coordinate (O1) at (rel axis cs:0,0);
            \coordinate (O2) at (rel axis cs:1,0);

            \addplot [
                bar shift=0pt,
                bar width=16pt,
                color=black!20, 
                fill=black!30,
                ultra thick,
                opacity=0.5,
] table [
                skip first n=1, 
                x index = {0}, 
                y index = {8}, 
                col sep=comma,
                each nth point=1,
            ]{./paperwithcode.csv};
            \addlegendentry{Total}

            \edef\cstmlst{
            "1.7\%",
            "1.3\%",
            "3.8\%",
            "4.1\%",
            "5.8\%",
            "4.9\%",
            }
            \addplot [
                bar shift=0pt,
                bar width=8pt,
                color=icra,
                fill=icra,
                ultra thick,
nodes near coords={\scriptsize \pgfmathsetmacro{\mystring}{{\cstmlst}[\coordindex]}\mystring},
            ] table [
                skip first n=1, 
                x index = {0}, 
                y index = {5}, 
                col sep=comma,
                each nth point=1,
            ]{./paperwithcode.csv};
            \addlegendentry{With Code}

            \draw [black!50,  thick ,sharp plot,dashed,] (A -| O1) -- (A -| O2) node[rotate=90,above,pos=0.99] {\scriptsize \hspace{4.5em} avg. 964};
            \draw [icra,  thick,sharp plot] (B -| O1) -- (B -| O2) node[rotate=90,above,pos=0.99] {\scriptsize \hspace{3.5em} avg. 38};

   \nextgroupplot[
        title={CDC},
        xtick={2016, 2017, 2018, 2019, 2020, 2021},
ybar, bar width=25pt,
        enlarge x limits=0.25,
ylabel near ticks,
        yticklabel=\empty, 
        ]

            \coordinate (A) at (axis cs:2016,1030);
            \coordinate (B) at (axis cs:2016,11);
            \coordinate (O1) at (rel axis cs:0,0);
            \coordinate (O2) at (rel axis cs:1,0);

            \addplot [
                bar shift=0pt,
                bar width=16pt,
                color=black!20, 
                fill=black!30,
                ultra thick,
                opacity=0.5,
] table [
                skip first n=1, 
                x index = {0}, 
                y index = {12}, 
                col sep=comma,
                each nth point=1,
            ]{./paperwithcode.csv};
            \addlegendentry{Total}

            \edef\cstmlst{
            "0.6\%",
            "0.5\%",
            "0.5\%",
            "1.2\%",
            "1.6\%",
            "2.6\%",
            }
            \addplot [
                bar shift=0pt,
                bar width=8pt,
                color=cdc,
                fill=cdc,
                ultra thick,
nodes near coords={\scriptsize \pgfmathsetmacro{\mystring}{{\cstmlst}[\coordindex]}\mystring},
            ] table [
                skip first n=1, 
                x index = {0}, 
                y index = {9}, 
                col sep=comma,
                each nth point=1,
            ]{./paperwithcode.csv};
            \addlegendentry{With Code}

            \draw [black!50,  thick ,sharp plot,dashed,] (A -| O1) -- (A -| O2) node[rotate=90,above,pos=0.99] {\scriptsize \hspace{4.5em} avg. 1030};
            \draw [cdc,  thick,sharp plot] (B -| O1) -- (B -| O2) node[rotate=90,above,pos=0.99] {\scriptsize \hspace{3.5em} avg. 11};

    \end{groupplot}
\end{tikzpicture}     \caption
    {
        The percentages of NeurIPS, ICRA, and CDC publications containing open-source code~(OSC) from 2016 to 2021.
        The percentage of publications with OSC has generally increased over the years~(except for NeurIPS from 2019 to 2020, ICRA from 2020 to 2021, and for CDC from 2016 to 2017). However, the magnitude of the percentages varies among the different conferences. 
        The start of the NeurIPS reproducibility program (dashed orange vertical line)~\cite{joelle2020}, which included encouragement of OSC submission in the call for papers, yields an exceptionally high 20.6\% increase between 2018 and 2019 for NeurIPS. 
        ICRA's and CDC's calls for papers do not encourage OSC explicitly~\cite{icraCfP2020, cdcCfP2021}.        
    }
    \label{fig:percent-papers-all}
\end{figure*}
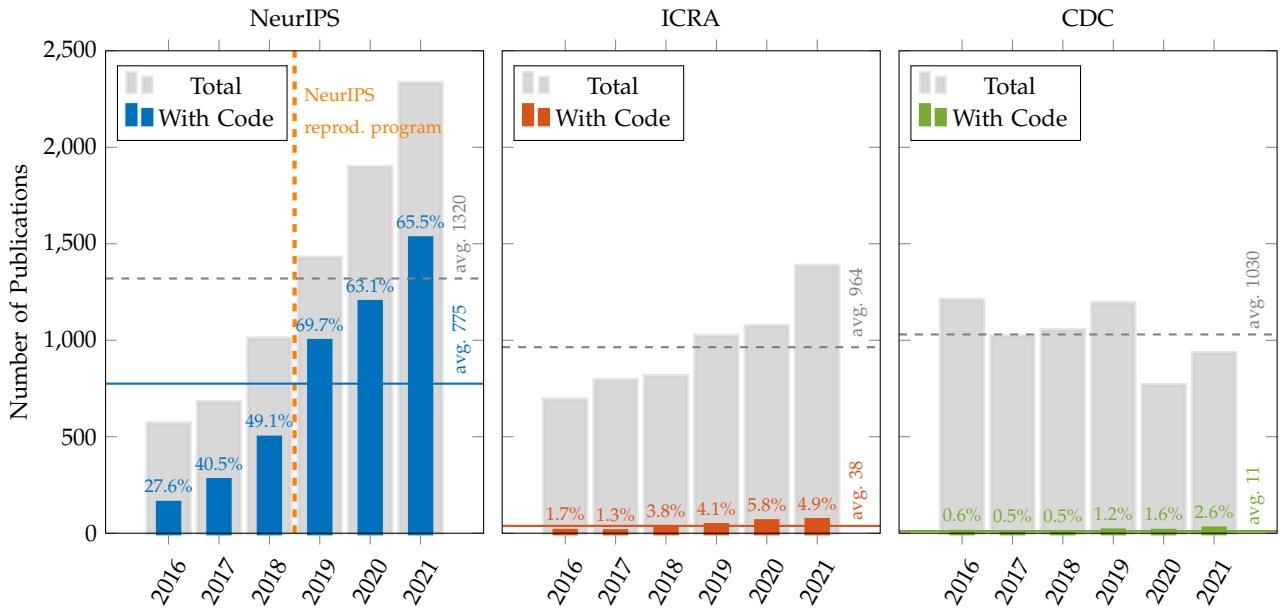

We give a more detailed overview of the percentages and total publication count for the three conferences in~\autoref{fig:percent-papers-all}. 
The percentage of NeurIPS papers containing OSC increased from 27.6\% in 2016 to more than 60\% from 2019 onward.   
NeurIPS
has recently made substantial efforts in improving the reproducibility of its publications~\cite{joelle2020}. 
We note that between 2018 and 2019, there is a 20.6\% increase in accepted publications with OSC. This jump correlates with the NeurIPS reproducibility checklist introduced in 2019 by~\cite{joelle2020}~(see dashed dark orange vertical line in~\autoref{fig:percent-papers-all} for NeurIPS), which advocated for submitting papers with OSC to allow others to better reproduce their results.
The correlation between the steep increase in publications with OSC and the beginning of the NeurIPS reproducibility program implies that this initiative was effective at increasing the percentage of publications with OSC at NeurIPS. 
Based on our collected data, we find that, from 2016 to 2020, the top $1\%$-cited publications at NeurIPS included OSC. 
ICRA has only achieved more than $5\%$ of publications with OSC once, and CDC only recently surpassed $2\%$ of publications with OSC in 2021 for the first time. 
Although the availability of OSC at ICRA and CDC is not as common, there is also a general upward trend of including OSC with publications for both conferences.
In contrast to NeurIPS, ICRA's and CDC's call for papers do not yet explicitly encourage the submission of OSC~\cite{icraCfP2020, cdcCfP2021}.

 \section{Scientific Impact of Open-Source Code}

In this section, we investigate the impact of~OSC 
for research publications using the collected publication data. 
To measure the academic impact of OSC in publications, we \textit{(i)} compare the number of citations for publications with OSC and without OSC and \textit{(ii)} investigate the correlation of a publication's OSC popularity with its number of citations. 

\begin{figure*}[tb!]
    \centering

    \begin{subfigure}[t]{0.32\linewidth}

\begin{tikzpicture}

\begin{axis}[
	width = 1.0\columnwidth, 
	height = 8cm,
    legend pos=north west,
	ylabel near ticks,
xlabel={Years since Pub. (from 2022)},
xmin=-0.5, xmax=5.5,
xtick={0,1,2,3,4,5},
xticklabels={1 (2021),2 (2020),3 (2019),4 (2018),5 (2017),6 (2016)},
x tick label style={rotate=60},
x label style={at={(axis description cs:0.5,-0.15)},anchor=north},
title={NeurIPS},
ylabel={Semantic Scholar Citations},
ymin=-67.8, ymax=1423.8,
]
\path [draw=black, fill=black!20, thick]
(axis cs:-0.396,1)
--(axis cs:-0.004,1)
--(axis cs:-0.004,8)
--(axis cs:-0.396,8)
--(axis cs:-0.396,1)
--cycle;
\path [draw=black, fill=neurips, thick]
(axis cs:0.004,2)
--(axis cs:0.396,2)
--(axis cs:0.396,9)
--(axis cs:0.004,9)
--(axis cs:0.004,2)
--cycle;
\path [draw=black, fill=black!20, thick]
(axis cs:0.604,5)
--(axis cs:0.996,5)
--(axis cs:0.996,20)
--(axis cs:0.604,20)
--(axis cs:0.604,5)
--cycle;
\path [draw=black, fill=neurips, thick]
(axis cs:1.004,8)
--(axis cs:1.396,8)
--(axis cs:1.396,37)
--(axis cs:1.004,37)
--(axis cs:1.004,8)
--cycle;
\path [draw=black, fill=black!20, thick]
(axis cs:1.604,7)
--(axis cs:1.996,7)
--(axis cs:1.996,37)
--(axis cs:1.604,37)
--(axis cs:1.604,7)
--cycle;
\path [draw=black, fill=neurips, thick]
(axis cs:2.004,14)
--(axis cs:2.396,14)
--(axis cs:2.396,75)
--(axis cs:2.004,75)
--(axis cs:2.004,14)
--cycle;
\path [draw=black, fill=black!20, thick]
(axis cs:2.604,10)
--(axis cs:2.996,10)
--(axis cs:2.996,53.25)
--(axis cs:2.604,53.25)
--(axis cs:2.604,10)
--cycle;
\path [draw=black, fill=neurips, thick]
(axis cs:3.004,23)
--(axis cs:3.396,23)
--(axis cs:3.396,154.25)
--(axis cs:3.004,154.25)
--(axis cs:3.004,23)
--cycle;
\path [draw=black, fill=black!20, thick]
(axis cs:3.604,11)
--(axis cs:3.996,11)
--(axis cs:3.996,58)
--(axis cs:3.604,58)
--(axis cs:3.604,11)
--cycle;
\path [draw=black, fill=neurips, thick]
(axis cs:4.004,45.75)
--(axis cs:4.396,45.75)
--(axis cs:4.396,346.25)
--(axis cs:4.004,346.25)
--(axis cs:4.004,45.75)
--cycle;
\path [draw=black, fill=black!20, thick]
(axis cs:4.604,15)
--(axis cs:4.996,15)
--(axis cs:4.996,67.75)
--(axis cs:4.604,67.75)
--(axis cs:4.604,15)
--cycle;
\path [draw=black, fill=neurips, thick]
(axis cs:5.004,41.75)
--(axis cs:5.396,41.75)
--(axis cs:5.396,620.25)
--(axis cs:5.004,620.25)
--(axis cs:5.004,41.75)
--cycle;
\draw[draw=black,fill=black!20,line width=0.3pt] (axis cs:0,0) rectangle (axis cs:0,0);
\addlegendimage{ybar,ybar legend,draw=black,fill=black!20,line width=0.3pt}
\addlegendentry{Without Code}

\draw[draw=black,fill=neurips,line width=0.3pt] (axis cs:0,0) rectangle (axis cs:0,0);
\addlegendimage{ybar,ybar legend,draw=black,fill=neurips,line width=0.3pt}
\addlegendentry{With Code}

\addplot [thick, black]
table {-0.200000047683716 1
-0.200000047683716 0
};
\addplot [thick, black]
table {-0.200000047683716 8
-0.200000047683716 18
};
\addplot [thick, black]
table {-0.297999978065491 0
-0.101999998092651 0
};
\addplot [thick, black]
table {-0.297999978065491 18
-0.101999998092651 18
};
\addplot [thick, black]
table {0.200000047683716 2
0.200000047683716 0
};
\addplot [thick, black]
table {0.200000047683716 9
0.200000047683716 19
};
\addplot [thick, black]
table {0.101999998092651 0
0.297999978065491 0
};
\addplot [thick, black]
table {0.101999998092651 19
0.297999978065491 19
};
\addplot [thick, black]
table {0.799999952316284 5
0.799999952316284 0
};
\addplot [thick, black]
table {0.799999952316284 20
0.799999952316284 42
};
\addplot [thick, black]
table {0.702000021934509 0
0.898000001907349 0
};
\addplot [thick, black]
table {0.702000021934509 42
0.898000001907349 42
};
\addplot [thick, black]
table {1.20000004768372 8
1.20000004768372 0
};
\addplot [thick, black]
table {1.20000004768372 37
1.20000004768372 80
};
\addplot [thick, black]
table {1.10199999809265 0
1.29799997806549 0
};
\addplot [thick, black]
table {1.10199999809265 80
1.29799997806549 80
};
\addplot [thick, black]
table {1.79999995231628 7
1.79999995231628 0
};
\addplot [thick, black]
table {1.79999995231628 37
1.79999995231628 82
};
\addplot [thick, black]
table {1.70200002193451 0
1.89800000190735 0
};
\addplot [thick, black]
table {1.70200002193451 82
1.89800000190735 82
};
\addplot [thick, black]
table {2.20000004768372 14
2.20000004768372 0
};
\addplot [thick, black]
table {2.20000004768372 75
2.20000004768372 166
};
\addplot [thick, black]
table {2.10199999809265 0
2.29800009727478 0
};
\addplot [thick, black]
table {2.10199999809265 166
2.29800009727478 166
};
\addplot [thick, black]
table {2.79999995231628 10
2.79999995231628 0
};
\addplot [thick, black]
table {2.79999995231628 53.25
2.79999995231628 118
};
\addplot [thick, black]
table {2.70199990272522 0
2.89800000190735 0
};
\addplot [thick, black]
table {2.70199990272522 118
2.89800000190735 118
};
\addplot [thick, black]
table {3.20000004768372 23
3.20000004768372 0
};
\addplot [thick, black]
table {3.20000004768372 154.25
3.20000004768372 348
};
\addplot [thick, black]
table {3.10199999809265 0
3.29800009727478 0
};
\addplot [thick, black]
table {3.10199999809265 348
3.29800009727478 348
};
\addplot [thick, black]
table {3.79999995231628 11
3.79999995231628 1
};
\addplot [thick, black]
table {3.79999995231628 58
3.79999995231628 128
};
\addplot [thick, black]
table {3.70199990272522 1
3.89800000190735 1
};
\addplot [thick, black]
table {3.70199990272522 128
3.89800000190735 128
};
\addplot [thick, black]
table {4.19999980926514 45.75
4.19999980926514 1
};
\addplot [thick, black]
table {4.19999980926514 346.25
4.19999980926514 792
};
\addplot [thick, black]
table {4.10200023651123 1
4.2979998588562 1
};
\addplot [thick, black]
table {4.10200023651123 792
4.2979998588562 792
};
\addplot [thick, black]
table {4.80000019073486 15
4.80000019073486 0
};
\addplot [thick, black]
table {4.80000019073486 67.75
4.80000019073486 146
};
\addplot [thick, black]
table {4.7020001411438 0
4.89799976348877 0
};
\addplot [thick, black]
table {4.7020001411438 146
4.89799976348877 146
};
\addplot [thick, black]
table {5.19999980926514 41.75
5.19999980926514 1
};
\addplot [thick, black]
table {5.19999980926514 620.25
5.19999980926514 1356
};
\addplot [thick, black]
table {5.10200023651123 1
5.2979998588562 1
};
\addplot [thick, black]
table {5.10200023651123 1356
5.2979998588562 1356
};
\addplot [thick, black]
table {-0.396000027656555 4
-0.00399994850158691 4
};
\addplot [thick, black]
table {0.00399994850158691 4
0.396000027656555 4
};
\addplot [thick, black]
table {0.603999972343445 11
0.996000051498413 11
};
\addplot [thick, black]
table {1.00399994850159 18
1.39600002765656 18
};
\addplot [thick, black]
table {1.60399997234344 16
1.99600005149841 16
};
\addplot [thick, black]
table {2.00399994850159 32
2.39599990844727 32
};
\addplot [thick, black]
table {2.60400009155273 26
2.99600005149841 26
};
\addplot [thick, black]
table {3.00399994850159 63
3.39599990844727 63
};
\addplot [thick, black]
table {3.60400009155273 26
3.99600005149841 26
};
\addplot [thick, black]
table {4.00400018692017 137
4.39599990844727 137
};
\addplot [thick, black]
table {4.60400009155273 28.5
4.99599981307983 28.5
};
\addplot [thick, black]
table {5.00400018692017 154.5
5.39599990844727 154.5
};
\end{axis}

\end{tikzpicture}
 \label{fig:impact-neurips}
      \end{subfigure}\hfill
    \begin{subfigure}[t]{0.32\linewidth}

\begin{tikzpicture}

\begin{axis}[
	width = 1.0\columnwidth, 
	height = 8cm,
    legend pos=north west,
xlabel={Years since Pub. (from 2022)},
xmin=-0.5, xmax=5.5,
xtick={0,1,2,3,4,5},
xticklabels={1 (2021),2 (2020),3 (2019),4 (2018),5 (2017),6 (2016)},
x tick label style={rotate=60},
x label style={at={(axis description cs:0.5,-0.15)},anchor=north},
title={ICRA},
ymin=-10.15, ymax=213.15,
]
\path [draw=black, fill=black!20, thick]
(axis cs:-0.396,0)
--(axis cs:-0.004,0)
--(axis cs:-0.004,4)
--(axis cs:-0.396,4)
--(axis cs:-0.396,0)
--cycle;
\path [draw=black, fill=icra, thick]
(axis cs:0.004,1)
--(axis cs:0.396,1)
--(axis cs:0.396,6.25)
--(axis cs:0.004,6.25)
--(axis cs:0.004,1)
--cycle;
\path [draw=black, fill=black!20, thick]
(axis cs:0.604,2)
--(axis cs:0.996,2)
--(axis cs:0.996,9)
--(axis cs:0.604,9)
--(axis cs:0.604,2)
--cycle;
\path [draw=black, fill=icra, thick]
(axis cs:1.004,4)
--(axis cs:1.396,4)
--(axis cs:1.396,20)
--(axis cs:1.004,20)
--(axis cs:1.004,4)
--cycle;
\path [draw=black, fill=black!20, thick]
(axis cs:1.604,3)
--(axis cs:1.996,3)
--(axis cs:1.996,15)
--(axis cs:1.604,15)
--(axis cs:1.604,3)
--cycle;
\path [draw=black, fill=icra, thick]
(axis cs:2.004,7)
--(axis cs:2.396,7)
--(axis cs:2.396,36)
--(axis cs:2.004,36)
--(axis cs:2.004,7)
--cycle;
\path [draw=black, fill=black!20, thick]
(axis cs:2.604,5)
--(axis cs:2.996,5)
--(axis cs:2.996,21)
--(axis cs:2.604,21)
--(axis cs:2.604,5)
--cycle;
\path [draw=black, fill=icra, thick]
(axis cs:3.004,13.5)
--(axis cs:3.396,13.5)
--(axis cs:3.396,110)
--(axis cs:3.004,110)
--(axis cs:3.004,13.5)
--cycle;
\path [draw=black, fill=black!20, thick]
(axis cs:3.604,5)
--(axis cs:3.996,5)
--(axis cs:3.996,26)
--(axis cs:3.604,26)
--(axis cs:3.604,5)
--cycle;
\path [draw=black, fill=icra, thick]
(axis cs:4.004,11.5)
--(axis cs:4.396,11.5)
--(axis cs:4.396,86.5)
--(axis cs:4.004,86.5)
--(axis cs:4.004,11.5)
--cycle;
\path [draw=black, fill=black!20, thick]
(axis cs:4.604,7)
--(axis cs:4.996,7)
--(axis cs:4.996,28)
--(axis cs:4.604,28)
--(axis cs:4.604,7)
--cycle;
\path [draw=black, fill=icra, thick]
(axis cs:5.004,22)
--(axis cs:5.396,22)
--(axis cs:5.396,80.5)
--(axis cs:5.004,80.5)
--(axis cs:5.004,22)
--cycle;
\draw[draw=black,fill=black!20,line width=0.3pt] (axis cs:0,0) rectangle (axis cs:0,0);
\addlegendimage{ybar,ybar legend,draw=black,fill=black!20,line width=0.3pt}
\addlegendentry{Without Code}

\draw[draw=black,fill=icra,line width=0.3pt] (axis cs:0,0) rectangle (axis cs:0,0);
\addlegendimage{ybar,ybar legend,draw=black,fill=icra,line width=0.3pt}
\addlegendentry{With Code}

\addplot [thick, black]
table {-0.200000047683716 0
-0.200000047683716 0
};
\addplot [thick, black]
table {-0.200000047683716 4
-0.200000047683716 10
};
\addplot [thick, black]
table {-0.297999978065491 0
-0.101999998092651 0
};
\addplot [thick, black]
table {-0.297999978065491 10
-0.101999998092651 10
};
\addplot [thick, black]
table {0.200000047683716 1
0.200000047683716 0
};
\addplot [thick, black]
table {0.200000047683716 6.25
0.200000047683716 12
};
\addplot [thick, black]
table {0.101999998092651 0
0.297999978065491 0
};
\addplot [thick, black]
table {0.101999998092651 12
0.297999978065491 12
};
\addplot [thick, black]
table {0.799999952316284 2
0.799999952316284 0
};
\addplot [thick, black]
table {0.799999952316284 9
0.799999952316284 19
};
\addplot [thick, black]
table {0.702000021934509 0
0.898000001907349 0
};
\addplot [thick, black]
table {0.702000021934509 19
0.898000001907349 19
};
\addplot [thick, black]
table {1.20000004768372 4
1.20000004768372 0
};
\addplot [thick, black]
table {1.20000004768372 20
1.20000004768372 43
};
\addplot [thick, black]
table {1.10199999809265 0
1.29799997806549 0
};
\addplot [thick, black]
table {1.10199999809265 43
1.29799997806549 43
};
\addplot [thick, black]
table {1.79999995231628 3
1.79999995231628 0
};
\addplot [thick, black]
table {1.79999995231628 15
1.79999995231628 33
};
\addplot [thick, black]
table {1.70200002193451 0
1.89800000190735 0
};
\addplot [thick, black]
table {1.70200002193451 33
1.89800000190735 33
};
\addplot [thick, black]
table {2.20000004768372 7
2.20000004768372 0
};
\addplot [thick, black]
table {2.20000004768372 36
2.20000004768372 70
};
\addplot [thick, black]
table {2.10199999809265 0
2.29800009727478 0
};
\addplot [thick, black]
table {2.10199999809265 70
2.29800009727478 70
};
\addplot [thick, black]
table {2.79999995231628 5
2.79999995231628 0
};
\addplot [thick, black]
table {2.79999995231628 21
2.79999995231628 45
};
\addplot [thick, black]
table {2.70199990272522 0
2.89800000190735 0
};
\addplot [thick, black]
table {2.70199990272522 45
2.89800000190735 45
};
\addplot [thick, black]
table {3.20000004768372 13.5
3.20000004768372 2
};
\addplot [thick, black]
table {3.20000004768372 110
3.20000004768372 203
};
\addplot [thick, black]
table {3.10199999809265 2
3.29800009727478 2
};
\addplot [thick, black]
table {3.10199999809265 203
3.29800009727478 203
};
\addplot [thick, black]
table {3.79999995231628 5
3.79999995231628 0
};
\addplot [thick, black]
table {3.79999995231628 26
3.79999995231628 57
};
\addplot [thick, black]
table {3.70199990272522 0
3.89800000190735 0
};
\addplot [thick, black]
table {3.70199990272522 57
3.89800000190735 57
};
\addplot [thick, black]
table {4.19999980926514 11.5
4.19999980926514 6
};
\addplot [thick, black]
table {4.19999980926514 86.5
4.19999980926514 186
};
\addplot [thick, black]
table {4.10200023651123 6
4.2979998588562 6
};
\addplot [thick, black]
table {4.10200023651123 186
4.2979998588562 186
};
\addplot [thick, black]
table {4.80000019073486 7
4.80000019073486 0
};
\addplot [thick, black]
table {4.80000019073486 28
4.80000019073486 59
};
\addplot [thick, black]
table {4.7020001411438 0
4.89799976348877 0
};
\addplot [thick, black]
table {4.7020001411438 59
4.89799976348877 59
};
\addplot [thick, black]
table {5.19999980926514 22
5.19999980926514 4
};
\addplot [thick, black]
table {5.19999980926514 80.5
5.19999980926514 122
};
\addplot [thick, black]
table {5.10200023651123 4
5.2979998588562 4
};
\addplot [thick, black]
table {5.10200023651123 122
5.2979998588562 122
};
\addplot [thick, black]
table {-0.396000027656555 1
-0.00399994850158691 1
};
\addplot [thick, black]
table {0.00399994850158691 3
0.396000027656555 3
};
\addplot [thick, black]
table {0.603999972343445 4
0.996000051498413 4
};
\addplot [thick, black]
table {1.00399994850159 11
1.39600002765656 11
};
\addplot [thick, black]
table {1.60399997234344 7
1.99600005149841 7
};
\addplot [thick, black]
table {2.00399994850159 13
2.39599990844727 13
};
\addplot [thick, black]
table {2.60400009155273 10
2.99600005149841 10
};
\addplot [thick, black]
table {3.00399994850159 32
3.39599990844727 32
};
\addplot [thick, black]
table {3.60400009155273 12
3.99600005149841 12
};
\addplot [thick, black]
table {4.00400018692017 55
4.39599990844727 55
};
\addplot [thick, black]
table {4.60400009155273 15
4.99599981307983 15
};
\addplot [thick, black]
table {5.00400018692017 38
5.39599990844727 38
};
\end{axis}

\end{tikzpicture}
 \label{fig:impact-icra}  
    \end{subfigure}
    \begin{subfigure}[t]{0.32\linewidth}

\begin{tikzpicture}

\begin{axis}[
	width = 1.0\columnwidth, 
	height = 8cm,
    legend pos=north west,
xlabel={Years since Pub. (from 2022)},
xmin=-0.5, xmax=5.5,
xtick={0,1,2,3,4,5},
xticklabels={1 (2021),2 (2020),3 (2019),4 (2018),5 (2017),6 (2016)},
x tick label style={rotate=60},
x label style={at={(axis description cs:0.5,-0.15)},anchor=north},
title={CDC},
ymin=-2.65, ymax=55.65,
]
\path [draw=black, fill=black!20, thick]
(axis cs:-0.396,0)
--(axis cs:-0.004,0)
--(axis cs:-0.004,1)
--(axis cs:-0.396,1)
--(axis cs:-0.396,0)
--cycle;
\path [draw=black, fill=cdc, thick]
(axis cs:0.004,0)
--(axis cs:0.396,0)
--(axis cs:0.396,3)
--(axis cs:0.004,3)
--(axis cs:0.004,0)
--cycle;
\path [draw=black, fill=black!20, thick]
(axis cs:0.604,0)
--(axis cs:0.996,0)
--(axis cs:0.996,3)
--(axis cs:0.604,3)
--(axis cs:0.604,0)
--cycle;
\path [draw=black, fill=cdc, thick]
(axis cs:1.004,0)
--(axis cs:1.396,0)
--(axis cs:1.396,12)
--(axis cs:1.004,12)
--(axis cs:1.004,0)
--cycle;
\path [draw=black, fill=black!20, thick]
(axis cs:1.604,1)
--(axis cs:1.996,1)
--(axis cs:1.996,5)
--(axis cs:1.604,5)
--(axis cs:1.604,1)
--cycle;
\path [draw=black, fill=cdc, thick]
(axis cs:2.004,1)
--(axis cs:2.396,1)
--(axis cs:2.396,4.5)
--(axis cs:2.004,4.5)
--(axis cs:2.004,1)
--cycle;
\path [draw=black, fill=black!20, thick]
(axis cs:2.604,1)
--(axis cs:2.996,1)
--(axis cs:2.996,7)
--(axis cs:2.604,7)
--(axis cs:2.604,1)
--cycle;
\path [draw=black, fill=cdc, thick]
(axis cs:3.004,0.25)
--(axis cs:3.396,0.25)
--(axis cs:3.396,27)
--(axis cs:3.004,27)
--(axis cs:3.004,0.25)
--cycle;
\path [draw=black, fill=black!20, thick]
(axis cs:3.604,2)
--(axis cs:3.996,2)
--(axis cs:3.996,8)
--(axis cs:3.604,8)
--(axis cs:3.604,2)
--cycle;
\path [draw=black, fill=cdc, thick]
(axis cs:4.004,9.5)
--(axis cs:4.396,9.5)
--(axis cs:4.396,27.5)
--(axis cs:4.004,27.5)
--(axis cs:4.004,9.5)
--cycle;
\path [draw=black, fill=black!20, thick]
(axis cs:4.604,2)
--(axis cs:4.996,2)
--(axis cs:4.996,10)
--(axis cs:4.604,10)
--(axis cs:4.604,2)
--cycle;
\path [draw=black, fill=cdc, thick]
(axis cs:5.004,2)
--(axis cs:5.396,2)
--(axis cs:5.396,11)
--(axis cs:5.004,11)
--(axis cs:5.004,2)
--cycle;
\draw[draw=black,fill=black!20,line width=0.3pt] (axis cs:0,0) rectangle (axis cs:0,0);
\addlegendimage{ybar,ybar legend,draw=black,fill=black!20,line width=0.3pt}
\addlegendentry{Without Code}

\draw[draw=black,fill=cdc,line width=0.3pt] (axis cs:0,0) rectangle (axis cs:0,0);
\addlegendimage{ybar,ybar legend,draw=black,fill=cdc,line width=0.3pt}
\addlegendentry{With Code}

\addplot [thick, black]
table {-0.200000047683716 0
-0.200000047683716 0
};
\addplot [thick, black]
table {-0.200000047683716 1
-0.200000047683716 2
};
\addplot [thick, black]
table {-0.297999978065491 0
-0.101999998092651 0
};
\addplot [thick, black]
table {-0.297999978065491 2
-0.101999998092651 2
};
\addplot [thick, black]
table {0.200000047683716 0
0.200000047683716 0
};
\addplot [thick, black]
table {0.200000047683716 3
0.200000047683716 7
};
\addplot [thick, black]
table {0.101999998092651 0
0.297999978065491 0
};
\addplot [thick, black]
table {0.101999998092651 7
0.297999978065491 7
};
\addplot [thick, black]
table {0.799999952316284 0
0.799999952316284 0
};
\addplot [thick, black]
table {0.799999952316284 3
0.799999952316284 7
};
\addplot [thick, black]
table {0.702000021934509 0
0.898000001907349 0
};
\addplot [thick, black]
table {0.702000021934509 7
0.898000001907349 7
};
\addplot [thick, black]
table {1.20000004768372 0
1.20000004768372 0
};
\addplot [thick, black]
table {1.20000004768372 12
1.20000004768372 12
};
\addplot [thick, black]
table {1.10199999809265 0
1.29799997806549 0
};
\addplot [thick, black]
table {1.10199999809265 12
1.29799997806549 12
};
\addplot [thick, black]
table {1.79999995231628 1
1.79999995231628 0
};
\addplot [thick, black]
table {1.79999995231628 5
1.79999995231628 11
};
\addplot [thick, black]
table {1.70200002193451 0
1.89800000190735 0
};
\addplot [thick, black]
table {1.70200002193451 11
1.89800000190735 11
};
\addplot [thick, black]
table {2.20000004768372 1
2.20000004768372 0
};
\addplot [thick, black]
table {2.20000004768372 4.5
2.20000004768372 8
};
\addplot [thick, black]
table {2.10199999809265 0
2.29800009727478 0
};
\addplot [thick, black]
table {2.10199999809265 8
2.29800009727478 8
};
\addplot [thick, black]
table {2.79999995231628 1
2.79999995231628 0
};
\addplot [thick, black]
table {2.79999995231628 7
2.79999995231628 16
};
\addplot [thick, black]
table {2.70199990272522 0
2.89800000190735 0
};
\addplot [thick, black]
table {2.70199990272522 16
2.89800000190735 16
};
\addplot [thick, black]
table {3.20000004768372 0.25
3.20000004768372 0
};
\addplot [thick, black]
table {3.20000004768372 27
3.20000004768372 53
};
\addplot [thick, black]
table {3.10199999809265 0
3.29800009727478 0
};
\addplot [thick, black]
table {3.10199999809265 53
3.29800009727478 53
};
\addplot [thick, black]
table {3.79999995231628 2
3.79999995231628 0
};
\addplot [thick, black]
table {3.79999995231628 8
3.79999995231628 17
};
\addplot [thick, black]
table {3.70199990272522 0
3.89800000190735 0
};
\addplot [thick, black]
table {3.70199990272522 17
3.89800000190735 17
};
\addplot [thick, black]
table {4.19999980926514 9.5
4.19999980926514 4
};
\addplot [thick, black]
table {4.19999980926514 27.5
4.19999980926514 30
};
\addplot [thick, black]
table {4.10200023651123 4
4.2979998588562 4
};
\addplot [thick, black]
table {4.10200023651123 30
4.2979998588562 30
};
\addplot [thick, black]
table {4.80000019073486 2
4.80000019073486 0
};
\addplot [thick, black]
table {4.80000019073486 10
4.80000019073486 22
};
\addplot [thick, black]
table {4.7020001411438 0
4.89799976348877 0
};
\addplot [thick, black]
table {4.7020001411438 22
4.89799976348877 22
};
\addplot [thick, black]
table {5.19999980926514 2
5.19999980926514 0
};
\addplot [thick, black]
table {5.19999980926514 11
5.19999980926514 11
};
\addplot [thick, black]
table {5.10200023651123 0
5.2979998588562 0
};
\addplot [thick, black]
table {5.10200023651123 11
5.2979998588562 11
};
\addplot [thick, black]
table {-0.396000027656555 0
-0.00399994850158691 0
};
\addplot [thick, black]
table {0.00399994850158691 1
0.396000027656555 1
};
\addplot [thick, black]
table {0.603999972343445 1
0.996000051498413 1
};
\addplot [thick, black]
table {1.00399994850159 6
1.39600002765656 6
};
\addplot [thick, black]
table {1.60399997234344 2
1.99600005149841 2
};
\addplot [thick, black]
table {2.00399994850159 2
2.39599990844727 2
};
\addplot [thick, black]
table {2.60400009155273 3
2.99600005149841 3
};
\addplot [thick, black]
table {3.00399994850159 11
3.39599990844727 11
};
\addplot [thick, black]
table {3.60400009155273 4
3.99600005149841 4
};
\addplot [thick, black]
table {4.00400018692017 15.5
4.39599990844727 15.5
};
\addplot [thick, black]
table {4.60400009155273 5
4.99599981307983 5
};
\addplot [thick, black]
table {5.00400018692017 6
5.39599990844727 6
};
\end{axis}

\end{tikzpicture}
 \label{fig:impact-cdc}  
    \end{subfigure}
    \caption
    {
        Box plots of citation counts for publications at NeurIPS, ICRA, and CDC published in the last six years~(2016 to 2021) with and without open-source code~(OSC)~(with different scales on the y-axis).  
        The median and third quartile numbers of citations for publications over the years since publication with OSC are typically greater than for publications without OSC. For NeurIPS the median and third quartile also increase at a much higher rate for publications with OSC. 
        Due to the small number of publications with open-source code for ICRA and CDC~(see~\autoref{fig:percent-papers-all}), it is hard to draw meaningful conclusions.
    }
    \label{fig:impact-all}
\end{figure*}
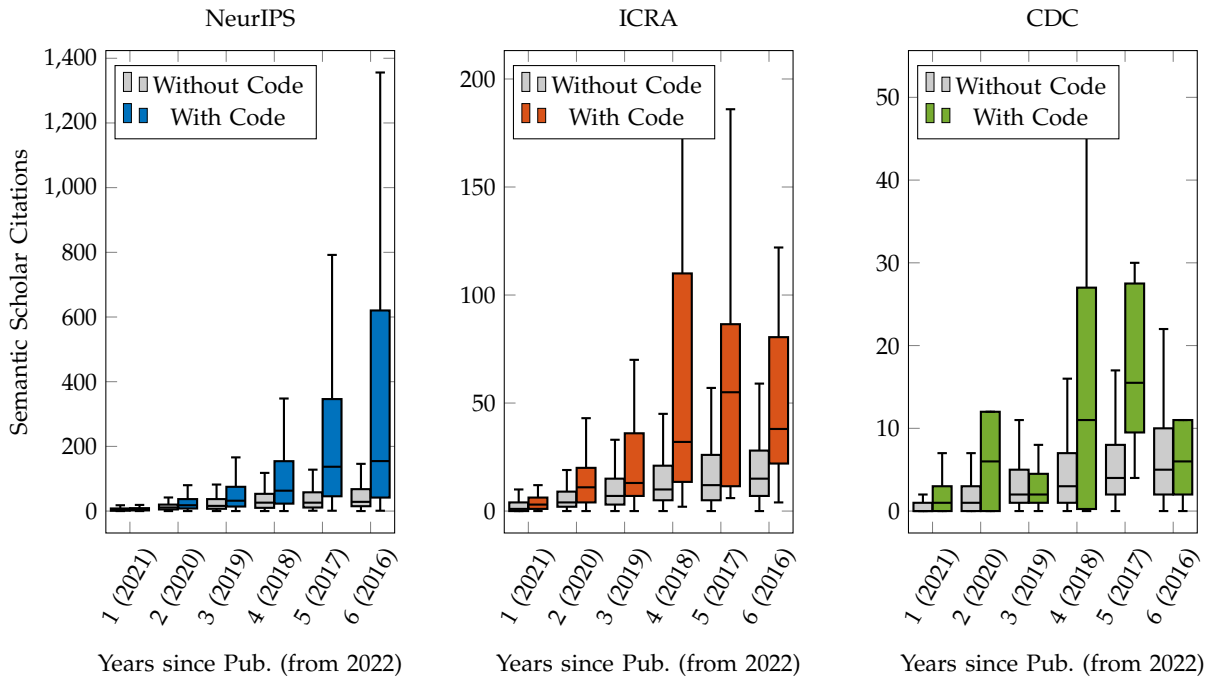

In~\autoref{fig:impact-all}, we present box plots~(outliers not shown) of publication citations at NeurIPS, ICRA, and CDC published from 2016 until 2021, with and without OSC. 
For NeurIPS, the median number of citations for publications with OSC is always greater or equal to the median number of citations for publications without OSC. 
We observe that \emph{publications with code get more citations over time than publications without code}.
Interestingly, we find that while the increase in the third quartile citation count for publications without OSC reduces to less than $20\%$ after four years since publication, the third quartile count for publications with OSC still increases by $79.13\%$. 
This implies that citation counts for highly cited publications with OSC at NeurIPS experience a greater growth rate even six years after publication.

In the last six years since publication, the median number of citations for publications with OSC at ICRA and CDC has typically been greater than the citation count for publications without OSC. 
This is also true for the third quartile of the citation count for most of the years measured.
Unlike the citation statistics for NeurIPS, there is no year-over-year monotonic increase in the median and the third quartile number of citations for publications with OSC at ICRA and CDC.   
We emphasize that the data for both ICRA and CDC for publications with OSC only rely on a few data points~(see~\autoref{fig:percent-papers-all}). Therefore, their significance might be limited.
However, despite the small number of publications with OSC at ICRA and CDC, the box plots suggest a positive impact on the citation count for publications with OSC.

\begin{figure}[tb!]
    \centering
\begin{tikzpicture}

\definecolor{darkturquoise0191191}{RGB}{0,191,191}
\definecolor{darkviolet1910191}{RGB}{191,0,191}
\definecolor{goldenrod1911910}{RGB}{191,191,0}
\definecolor{green01270}{RGB}{0,127,0}

\begin{axis}[
width = 1.0\columnwidth, 
height = 9cm,
legend pos=south east,
xlabel={Github Stars},
xmin=-0.2, xmax=500,
xtick = {0, 100, 200, ..., 500},
ylabel={Semantic Scholar Citations},
ymin=-0.2, ymax=500,
axis equal image, 
]

\addlegendimage{empty legend}

\addplot [thick, blue]
table {-36.9009521525901 -213.58138758212
-43.3504049055584 -209.684818820858
-48.6879164084813 -200.862824504745
-52.8258448700441 -187.260261536554
-55.6962456193072 -169.100483624159
-57.2519867553244 -146.68167381883
-57.4675230516528 -120.371948357766
-56.3393154082657 -90.6033122056675
-53.8858889634765 -57.8645655451536
-50.1475289116799 -22.6932776912782
-45.185619021574 14.3330397817608
-39.0816337163748 52.6064157647039
-31.93580026604 91.4984024820356
-23.8654530582669 130.370394587084
-15.0031069710902 168.58411504416
-5.49428148224863 205.512095619832
4.50488875661669 240.547979893977
14.8302176466256 273.116479615433
25.3121635728632 302.682820919179
35.7786132730098 328.761525298412
46.0577079323496 350.924381148211
55.9806651007571 368.807474988247
65.3845500958621 382.117166912071
74.1149513860293 390.63491214626
82.0285160234392 394.220849549555
88.9953034955277 392.816098128932
94.9009193446106 386.443723863882
99.6483935217153 375.208360963662
103.159772632105 359.294493776477
105.377399927864 338.963427561576
106.264862030106 314.548997864126
105.807586835647 286.452088944787
104.013082790538 255.134051271231
100.910815601588 221.109126156226
96.5517244102734 184.93600192949
91.0073853734412 147.208640290765
84.3688363847949 108.546523475131
76.7450822351955 69.5844823718279
68.2613047570094 30.9622726186089
59.0568073418816 -6.68593016814738
49.2827275828966 -42.7419435395226
39.0995555994778 -76.6137287527817
28.6744987940639 -107.745112031159
18.1787363111838 -135.624916921896
7.78460828063503 -159.795357799467
-2.33721400268168 -179.859556692889
-12.0205304967074 -195.488060011037
-21.1063414123197 -206.424248161501
-29.4454579854753 -212.488549237198
-36.9009521525901 -213.58138758212
};
\addplot [thick, blue, forget plot]
table {85.6377942578533 394.142791090892
-36.9009521525901 -213.58138758212
};
\addplot [thick, blue, mark=x, mark size=3, mark options={solid}, only marks, forget plot]
table {24.3684210526316 90.280701754386
};

\addplot [thick, green01270]
table {-86.0190595284603 -197.298257953074
-101.35344157455 -187.269695181657
-114.05691435786 -172.062024371843
-123.920887206492 -151.924955057891
-130.783393958979 -127.189137361217
-134.531752444167 -98.2607327245407
-135.10441471936 -65.614744752228
-132.491977685829 -29.7872196642373
-126.737337487442 8.63355556799218
-117.934985157161 49.0167128773468
-106.229455076893 90.6991619372596
-91.8129517269415 132.99647808654
-74.9221936937455 175.21414056339
-55.8345267572402 216.65893651711
-34.8633698808161 256.650343543942
-12.353068880605 294.531703846128
11.3267577233003 329.681006534794
35.7872878327279 361.521101031545
60.626880219529 389.529173865013
85.4376694634396 413.245333253597
109.812263091006 432.280160515805
133.350430948736 446.321104314237
155.665676971154 455.137612739907
176.391585435508 458.584918968209
195.187837497814 456.606418326183
211.745799218859 449.23459773967
225.793589324952 436.590502298896
237.100543490924 418.881747701435
245.481001841967 396.399111208191
250.797357483648 369.511757088818
252.962316003308 338.661174954614
251.940328841908 304.353930511313
247.748177000384 267.153347764199
240.454695496076 227.670259253514
230.179643093645 186.552976201035
217.09173586947 144.476643257733
201.405876898336 102.132152647276
183.379627550892 60.2147997348109
163.30897834306 19.4128662962922
141.523488779948 -19.6036810492502
118.380875997892 -56.1941916730726
94.2611410589598 -89.7578504023316
69.5603293440233 -119.74354288861
44.684027498551 -145.658904890282
20.0407037111027 -167.07840687975
-3.96499932303274 -183.650341226519
-26.9389086237087 -195.102597226774
-48.5037932353558 -201.247129153602
-68.3055583488444 -201.983043962642
-86.0190595284603 -197.298257953074
};
\addplot [thick, green01270, forget plot]
table {203.764514073915 453.589167043983
-86.0190595284603 -197.298257953074
};
\addplot [thick, green01270, mark=x, mark size=3, mark options={solid}, only marks, forget plot]
table {58.8727272727273 128.145454545455
};

\addplot [thick, red]
table {-11.955377963892 -68.9461153092584
-18.4566514641321 -66.0617536608996
-24.3081111234814 -61.321931603027
-29.4136761361055 -54.8044768827912
-33.6895132656483 -46.6164059265138
-37.0654133846545 -36.8921666329254
-39.4859443065019 -25.7914307423857
-40.9113609803728 -13.4964720313208
-41.31825810389 -0.209173381854065
-40.6999544375349 13.8522881295376
-39.0666025104148 28.4570236935871
-36.4450219160167 43.3652239561025
-32.8782589352232 58.3320966809328
-28.4248797175703 73.1118862317177
-23.1580086266916 87.4619088726428
-17.1641275402888 101.146537628672
-10.5416558200909 113.941071273963
-3.39933426865535 125.635423919784
4.14556039160334 136.037574618998
11.9691411989536 144.976720344709
19.9429451728333 152.306080571234
27.9360426724392 157.905307406287
35.8171872578086 161.682461700034
43.456970752736 163.575522683397
50.7299481234477 163.553406347366
57.516697282589 161.616475841566
63.7057799965967 157.796535511355
69.1955716984131 152.156308671345
73.895930160068 144.788407690318
77.7296756255692 135.813813298686
80.6338581003631 125.379888088242
82.5607909885038 113.657956822481
83.478834105241 100.840493288716
83.3729132079786 87.1379598838145
82.2447675149214 72.7753518274573
80.11292114716 57.9885027468541
77.0123789631149 43.0202122951041
72.9940517797292 28.1162593876041
68.1239204182613 13.5213665190533
62.4819523010245 -0.524818573144287
56.1607883885233 -13.7916579181024
49.2642220174427 -26.0613103805587
41.9054946169302 -37.1323086062922
34.2054362874592 -46.8228671105107
26.2904817739199 -54.973867190485
18.290594410604 -61.4514696502484
10.3371321268482 -66.1493124364945
2.56069055346657 -68.9902571005521
-4.91104135389741 -69.9276554095796
-11.9553779638919 -68.9461153092584
};
\addplot [thick, red, forget plot]
table {54.1912753997894 162.823038386181
-11.955377963892 -68.9461153092584
};
\addplot [thick, red, mark=x, mark size=3, mark options={solid}, only marks, forget plot]
table {21.1179487179487 46.9384615384615
};
\addplot [thick, darkturquoise0191191]
table {-23.3819158516839 -33.8030570143246
-27.7042601076697 -30.1869502363641
-31.2740326889753 -25.6063015063217
-34.0326180285073 -20.1363249497058
-35.9347202312558 -13.8668374303502
-36.9491068316381 -6.90078375999363
-37.0591216297202 0.647453652683776
-36.2629581855592 8.65393295804809
-34.5736894808969 16.9871879872563
-32.0190532611617 25.5103869231141
-28.6409965824548 34.0835790747312
-24.4949870420491 42.5659928639968
-19.6491020019756 50.8183472910001
-14.1829107606205 58.7051389241984
-8.18616802704582 66.0968668630166
-1.757340151148 72.8721591390548
4.9980116910853 78.9197656404535
11.9689648091664 84.1403848356548
19.0410563456187 88.4482943018052
26.0981627529323 91.7727582845693
33.0244065390033 94.0591891772597
39.7060589723508 95.2700438478303
46.0334075047435 95.3854400960662
51.9025572482084 94.4034831187618
57.2171369262192 92.3402966223365
61.8898812873926 89.229758072022
65.844063998503 85.1229424248188
69.0147574887533 80.087283480115
71.3498990586921 74.2054666185643
72.8111457482996 67.5740711104218
73.3745039273306 60.3019842865964
73.0307232700604 52.5086136116875
71.7854486453771 44.3219260167179
69.6591274281904 35.8763466856637
66.6866737541012 27.3105517976502
62.9168952302602 18.7651914679244
58.4116915158078 10.3805802768512
53.245037931182 2.29439330838284
47.5017707854026 -5.36059447101304
41.2761943662222 -12.4586883736324
34.6705324663283 -18.8833378938316
27.7932498714937 -24.529050464108
20.7572713717931 -29.3031236401417
13.6781275396712 -33.1271672723688
6.67205772113806 -35.9383906700694
-0.145898611079229 -37.6906336227737
-6.66379080249833 -38.3551243496417
-12.7745952435329 -37.9209519313091
-18.3779726942035 -36.3952454668816
-23.3819158516839 -33.8030570143246
};
\addplot [thick, darkturquoise0191191, forget plot]
table {59.6387965856289 90.9131487574438
-23.3819158516839 -33.8030570143246
};
\addplot [thick, darkturquoise0191191, mark=x, mark size=3, mark options={solid}, only marks, forget plot]
table {18.1284403669725 28.5550458715596
};
\addplot [thick, darkviolet1910191]
table {52.5939710646118 44.8767745528854
50.0952634016943 47.4932212371035
47.032579346856 49.6004226887701
43.4562080876049 51.1637787183054
39.4248735411437 52.1576190629258
35.0047701092732 52.565624891649
30.2684757681443 52.381096760047
25.2937603398963 51.6070646149393
20.1623085143277 50.2562380427494
14.9583785885477 48.3507975784469
9.76741894806719 45.9220305017769
4.67466500667099 43.0098171009914
-0.236260356719919 39.661975839619
-4.88471988340853 35.9334781786152
-9.19438599648021 31.8855459464938
-13.0944940975055 27.5846460785952
-16.5210045192898 23.1013992318455
-19.4176540554385 18.5094201955221
-21.7368798006501 13.8841091384496
-23.4406001323069 9.30141354031126
-24.5008400096648 4.83658113613318
-24.9001903232495 0.562924350554965
-24.6320937519527 -3.44938349015579
-23.7009524340631 -7.13446040218085
-22.1220556842776 -10.4317975245042
-19.9213289435768 -13.2872526727372
-17.1349080842006 -15.653939352289
-13.8085460596252 -17.4929966333121
-9.99686164233188 -18.7742272460955
-5.7624425850729 -19.4765934193989
-1.17481793170161 -19.5885623200745
3.69068364780535 -19.1082954218689
8.75417075202653 -18.0436786939728
13.9325010626983 -16.4121931136255
19.140646539282 -14.2406276289513
24.2930895770393 -11.5646392851724
29.3052272037369 -8.42816773691208
34.0947602581127 -4.88271376028021
38.5830447398275 -0.986493611553783
42.696383141777 3.19651688214198
46.3672345611204 7.5976328041461
49.5353237190383 12.1445879526948
52.1486306791443 16.7627214499561
54.1642450134022 21.3762036715803
55.5490703901682 25.9092813670349
56.2803680150412 30.2875215258841
56.3461300012388 34.4390335657628
55.7452765387656 38.2956497737508
54.4876736248639 41.7940446183349
52.5939710646118 44.8767745528853
};
\addplot [thick, darkviolet1910191, forget plot]
table {-21.097395722146 -11.9178704432963
52.5939710646118 44.8767745528854
};
\addplot [thick, darkviolet1910191, mark=x, mark size=3, mark options={solid}, only marks, forget plot]
table {15.7482876712329 16.4794520547945
};
\addplot [thick, goldenrod1911910]
table {28.613548080188 6.7284290951466
28.2757607933381 7.9696009149451
27.5982943659035 9.14527048751756
26.592272777893 10.2361333509855
25.2742148759688 11.2242775672892
23.6657631344409 12.0934778357159
21.7933282864371 12.8294619118882
19.6876556603954 13.4201449576227
17.3833203426208 13.8558279736388
14.9181594553245 14.1293570568587
12.3326508721228 14.236240867298
9.6692485724717 14.1747243757443
6.97168554849766 13.945817681288
4.28425571047622 13.5532794255238
1.65108658204966 13.003555075759
-0.884585272491119 12.3056710906188
-3.28112419139701 11.4710867058451
-5.49917907187547 10.5135057739625
-7.50232951411047 9.44865174739895
-9.2576838432206 8.29400949983101
-10.7364191896513 7.06853822504195
-11.9142547598959 5.79236012748385
-12.7718505264542 4.48643001623675
-13.2951247905486 3.17219122762294
-13.4754854032296 1.87122352621738
-13.3099708482231 0.60488876570995
-12.8012988699466 -0.606019872141538
-11.9578218482202 -1.74161930098093
-10.7933896524217 -2.78326300958204
-9.32712222701349 -3.71384723704742
-7.58309564258366 -4.51809181608947
-5.58994676743233 -5.18279107295488
-3.38040305098591 -5.69703066422251
-0.990745139982305 -6.05236679002067
1.53978884874543 -6.24296484098755
4.16964761706281 -6.26569520239564
6.85564895670041 -6.12018464233904
9.55368879991992 -5.80882244019102
12.219465407087 -5.33672115470278
14.8092068000113 -4.71163267593008
17.2803894966183 -3.94382093941289
19.5924367453493 -3.04589339263749
21.7073847943001 -2.0325939810974
23.5905062549778 -0.920561053115946
25.2108803250597 0.271945841373251
26.5419005071129 1.52534577213129
27.5617114865459 2.81905794694725
28.2535679952709 4.13183964767724
28.6061097685417 5.4421350350891
28.613548080188 6.7284290951466
};
\addplot [thick, goldenrod1911910, forget plot]
table {-13.4359257025656 1.23241006569256
28.613548080188 6.7284290951466
};
\addplot [thick, goldenrod1911910, mark=x, mark size=3, mark options={solid}, only marks, forget plot]
table {7.58881118881119 3.98041958041958
};

\legend{
    \hspace{-.6cm}Pub. Year,
    2016,
    2017,
    2018,
    2019,
    2020,
    2021
}

\end{axis}

\end{tikzpicture}
     \caption
    {   
        The uncertainty ellipses with confidence value $p = 0.99$ of
        the number of Semantic Scholar citations for a NeurIPS publication published in the last six years (2016 - 2021) over the corresponding GitHub repository's number of stars.
        The uncertainty ellipses' area and the steepness of its major principal axis almost monotonically increase with additional years since publication. 
        This highlights that publications with highly popular open-source code tend to be cited more often.  
}
    \label{fig:stars-vs-citations-neurips}
\end{figure}
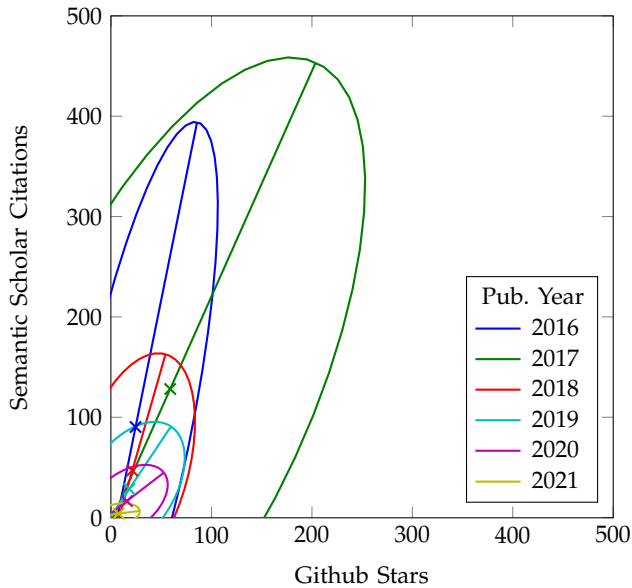

We also investigated the correlation between a repository's popularity and the associated publication's number of Semantic Scholar citations. A proxy for a repository's popularity is the number of GitHub stars or forks~\cite{Borges2016, Borges2018}.
As the number of stars is typically strongly correlated with the number of forks~\cite{Borges2016}, we chose to only focus on GitHub stars for simplicity.
In \autoref{fig:stars-vs-citations-neurips}, the number of Semantic Scholar citations for a NeurIPS publication is plotted against the number of GitHub stars, for publications published up to six years ago are shown. We summarized the data using uncertainty ellipses with a confidence value $p = 0.99$. We used the minimum covariance determinant algorithm by Rousseeuw and van~Driessen~(1999) to robustly fit the uncertainty ellipses despite outliers~\cite{Rousseeuw1999}. 
This method assumes an underlying unimodal distribution. 
Under this assumption, we see that the ellipses' major principal axes for each uncertainty ellipse have a positive slope. This implies there is a positive correlation between the popularity of OSC and its associated publication's citation count.
Furthermore, we find that this correlation typically increases with additional years since publication.
This positive correlation suggests that \emph{
highly-popular OSC associated with a publication could potentially be one of the factors that increase a publication's citation count and impact}. 

To summarize, the results in this section highlight the positive impact of OSC and OSC popularity on a publication's citation count---a common measure to assess academic impact.

 \section{Conclusion and Lessons Learned}
\label{sec:conclusion}
In this article, we investigated the current status and the impact of OSC at premier conferences in three research fields. 
We found that 60\% of the publications at the machine learning conference~(NeurIPS)
included OSC since 2019.
One core factor contributing to these results was the introduction of the reproducibility program at NeurIPS in 2019.  
We also determined that there is an evident positive trend at robotics and control conferences~(ICRA and CDC): the percentages of publications with OSC have at least tripled since 2016 and increased to almost 5.0\% and 2.6\%, respectively. 
While the existence of OSC improves reproducibility, we also find that it positively correlates with increased academic impact (as measured by the citation count). 
Furthermore, the GitHub repository's popularity (as measured by stars) also positively correlates with increased academic impact and typically increases with the number of years since publication.

We believe that \emph{robotic and control theory conferences like ICRA and CDC would benefit from more publications with OSC}. 
Some obstacles that keep researchers from 
releasing their implementations as OSC are the additional required effort and risk of releasing erroneous OSC~\cite{the_turing_way2022}. 
In light of these barriers, the OSC submission policy has been softly enforced at NeurIPS~\cite{neuripsCfP2021}. 
This still retains the flexibility
for researchers who cannot submit implementations, while encouraging OSC.
In addition to the positive academic impacts of releasing OSC with publications determined in this article, the authors in~\cite{the_turing_way2022} highlight further positive aspects of reproducible research and releasing OSC (e.g. the motivation to conduct analyses at the highest standard and the improved reusability of implementations for the authors themselves).

In the spirit of this article, we published the data and the OSC that we used to 
analyze it~\cite{osc-repo}. 
In the future, we plan to include data from additional years, conferences, journals, and additional code hosting platforms like GitLab and Bitbucket.  

Finally, we end with a list of lessons learned for improving the reproducibility of scientific work, 
especially at conferences in robotics and control theory: 
\begin{enumerate}
\item Encourage the submission of OSC: 
    This can improve reproducibility and the impact of the publication as discussed in this article. 
    This could be implemented by adding statements around OSC submission in the call for papers and a submission checklist for authors as presented in~\cite{joelle2020}.
    We also suggest that paper submission websites contain specific fields for linking open-source code and datasets. 
\item Reproducibility challenges:
As proposed by~\cite{joelle2020} for machine learning, these challenges could also be part of graduate courses in control theory~(also suggested by~\cite{how2018control}) or robotics. 
Such challenges will provide insight into the reproducibility of 
publications and increase awareness around open research. 
\item Unified interfaces: Another channel to promote OSC is to encourage the reusability of the code through unified interfaces. ROS and OpenAI Gym are examples that facilitated code sharing and algorithms comparisons within the respective communities. One valuable next step would be further encouraging frameworks with structured and standardized interfaces to lower the barrier for cross-discipline adoptions.
\item Releasing benchmarks and hosting competitions: Competitions have been growing more popular at NeurIPS and ICRA. However, at CDC they are only hosted for the first time in 2023. 
Benchmarks and competitions are great ways to drive novel research ideas and reduce the gap between academic research and real-world application. Defining the right challenges~\cite{stanley, nimbro, cerberus, zeus}, however, would require closer interactions between academia and industry. 
    
\end{enumerate}

\section{Appendix A: Conference and Citation Data Collection}
\labelname{Appendix A}
\label{sec:appendix}
To systematically collect data and statistics for CDC, ICRA, and NeurIPS, we developed an ad-hoc methodology to suit the different platforms the papers were hosted on---being CDC and ICRA papers on IEEE Xplore, and NeurIPS papers on \texttt{papers.nips.cc}.

\subsection{CDC and ICRA Data Collection Procedure}

The conference name, conference year, publication title, first and last author affiliations, keywords, benchmarks, experimental results, and potential links to OSC were obtained by scraping the HTML from the IEEE Xplore page for each paper. 
We note that not all conference publications are available on IEEE Xplore. This leads to a discrepancy between the numbers reported by the conference, e.g.~\cite{icra-num-submissions2, cdc-2021-final-program} and the presented numbers in~\autoref{fig:percent-papers-all}. 
Should a GitHub link be found, it was then scraped for the number of stars associated with that repository. Finally, citations were scraped by querying the paper title on Semantic Scholar. Data for CDC was collected between August 7--9th, 2022, and data for ICRA was collected between August 23rd to September 1st, 2022.

\subsection{NeurIPS Data Collection Procedure}

The conference name, conference year, publication title, author names, keywords, and benchmarks were scraped from the publications in the NeurIPS proceedings. Potential links to OSC were scraped by first performing a regex search for \texttt{github.com} and \texttt{github.io} links in the paper. If nothing was found, the paper title was further queried on \texttt{paperswithcode.com}. Finally, for publications from 2021, Openreview.net explicitly lists any OSC released along the paper in a designated field. This information is always used for publications from 2021 specifically. Should a GitHub link be found, its associated stars and forks were scraped. Finally, citation data were scraped from querying the paper title on Semantic Scholar. All NeurIPS data were collected between August 8th to 9th, 2022. 

\subsection{Data Collection Accuracy}                   

Because the data has been collected with automated scraping and processing tools, there are no guarantees that some links to OSC have not been missed or that a publication has not been matched with an incorrect entry on Semantic Scholar. 
Our implementation mainly focused on OSC hosted on GitHub. Therefore, OSC hosted on other platforms or personal servers could be absent in the statistics provided in this work.

To quantify the accuracy of our method, we compare the percentages provided in \autoref{fig:percent-papers-all} with the percentages of camera-ready papers with available OSC provided by~\cite{joelle2020}. For NeurIPS in 2018, the authors in~\cite{joelle2020} state $<50\%$, while we provide $49.1\%$. 
For NeurIPS in 2019, the authors in~\cite{joelle2020} list $74.4\%$ compared to $69.7\%$ shown in \autoref{fig:percent-papers-all}. This yields an error of $4.7\%$. The data for~OSC for NeurIPS in 2021 was directly pulled from Openreview.net as mentioned above. To the best knowledge of the authors, no other sources state statistics for these quantities. Therefore, the maximum quantifiable error is $4.7\%$, which is a sizable number of possible implementations not accounted for. However, we believe that 
these deviations in the statistics generated from our collected data are not significant enough to invalidate the overall discussion and conclusions in this article. 
 
\section{Acknowledgement}
The authors would like to acknowledge the early contributions to this work by Richard Hanxu.

\clearpage
\section{Author Information}

\begin{IEEEbiography}{{S}iqi Zhou}{\,}(siqi.zhou@tum.de) received the B.A.Sc. degree (Engineering Science, 2016) and the Ph.D. degree (Aerospace Engineering, 2022) from the University of Toronto. She was a postdoctoral fellow at the Vector Institute for Artificial Intelligence and recently joined the Learning Systems and Robotics lab at the Technical University of Munich. Her research combines control theory and machine learning and is centred on safe and deployable decision-making for robotic systems. \end{IEEEbiography}

\begin{IEEEbiography}{{L}ukas Brunke}{\,}obtained the B.Sc. degree (mechanical engineering, 2015) and the M.Sc. degree (mechatronics, 2018) from the Technical University of Hamburg. In 2018 he held research positions at UC Berkeley and the Max Planck Institute for Intelligent Systems, Tübingen. From 2019 to early 2020, he was part of the machine learning team at the Volkswagen Group Research Department in Silicon Valley. Since 2020 he has been pursuing the PhD degree at the University of Toronto under the supervision of Professor Angela Schoellig. Currently, he is a researcher at the Learning Systems and Robotics Lab at the Technical University of Munich. His research combines model-based control algorithms with machine learning methods for safe, high-performance robotic applications in uncertain and dynamic environments.
\end{IEEEbiography}

\begin{IEEEbiography}{{A}llen Tao}{\,}is an undergraduate student in the Engineering Science program at the University of Toronto. He is pursuing a major in Robotics Engineering with a minor in Artificial Intelligence. He was a summer research intern at the Learning Systems and Robotics Lab under Professor Angela Schoellig in 2022 and is currently working on indoor navigation at the Autonomous Space Robotics Laboratory led by Professor Tim Barfoot at the University of Toronto Institute for Aerospace Studies.

\end{IEEEbiography}

\begin{IEEEbiography}{{A}dam W. Hall} is with the Learning Systems and Robotics Lab at the University of Toronto Institute for Aerospace Studies, under the supervision of Professor Angela Schoellig. He is currently exploring how to combine machine learning with classical control theory to help robots safely improve their performance over time. His prior work also includes developing open-source benchmarking and reproducibility tools including the safe-control-gym.
\end{IEEEbiography}

\begin{IEEEbiography}{{F}ederico Pizarro Bejarano}{\,}received the B.A.Sc. degree (Engineering Science, 2021) from the University of Toronto. Since 2021 they have been pursuing the Ph.D. degree at the University of Toronto under the supervision of Professor Angela Schoellig. Currently, they are a researcher at the Learning Systems and Robotics Lab at the University of Toronto. Their research augments learning-based controllers with model-based filters that guarantee safety.
\end{IEEEbiography}

\begin{IEEEbiography}{{J}acopo Panerati}{\,} is the lead researcher in machine learning for control at the Technology Innovation Institute (Abu Dhabi, UAE). Their research interests include reinforcement learning, aerial robotics, distributed and multi-robot systems, probabilistic graphical models, software engineering, and embedded computing.
They hold a Ph.D. degree in computer engineering from Polytechnique Montr\'eal and received the M.Sc. degree in computer science from the University of Illinois at Chicago in 2012, the \emph{Laurea Triennale} degree in computer engineering from Politecnico di Milano in 2009, and the \emph{Laurea Specialistica} degree in computer engineering again from Politecnico di Milano in 2011.
In 2019, they were a visiting postdoctoral fellow at the European Astronaut Centre (K\"oln, Germany) and worked as a research associate in the University of Cambridge’s Department of Computer Science and Technology. From 2020 to 2022, they were a postdoctoral fellow at the University of Toronto Institute for Aerospace Studies (UTIAS) working with Professor~Angela~P. Schoellig in the Dynamic Systems Lab. 

\end{IEEEbiography}

\begin{IEEEbiography}{{A}ngela P. Schoellig}{\,}is an Alexander von Humboldt Professor for Robotics and Artificial Intelligence at the Technical University of Munich. She is also an Associate Professor at the University of Toronto Institute for Aerospace Studies and a Faculty Member of the Vector Institute in Toronto. She conducts research at the intersection of robotics, controls, and machine learning. Her goal is to enhance robots' performance, safety, and autonomy by enabling them to learn from past experiments and each other. In Canada, she has held a Canada Research Chair (Tier 2) in Machine Learning for Robotics and Control and a Canada CIFAR Chair in Artificial Intelligence and has been a principal investigator of the NSERC Canadian Robotics Network. She is a recipient of the Robotics: Science and Systems Early Career Spotlight Award (2019), a Sloan Research Fellowship (2017), and an Ontario Early Researcher Award (2017). 
Her team is the five-time winner of the North-American SAE AutoDrive Challenge (2018-21). Her Ph.D. at ETH Zurich (2013) was awarded the ETH Medal and the Dimitris N. Chorafas Foundation Award. She holds both an M.Sc. in Engineering Cybernetics from the University of Stuttgart (2008) and an M.Sc. in Engineering Science and Mechanics from the Georgia Institute of Technology (2007).
\end{IEEEbiography}


\begin{thebibliography}{10}
\providecommand{\url}[1]{#1}
\csname url@samestyle\endcsname
\providecommand{\newblock}{\relax}
\providecommand{\bibinfo}[2]{#2}
\providecommand{\BIBentrySTDinterwordspacing}{\spaceskip=0pt\relax}
\providecommand{\BIBentryALTinterwordstretchfactor}{4}
\providecommand{\BIBentryALTinterwordspacing}{\spaceskip=\fontdimen2\font plus
\BIBentryALTinterwordstretchfactor\fontdimen3\font minus
  \fontdimen4\font\relax}
\providecommand{\BIBforeignlanguage}[2]{{\expandafter\ifx\csname l@#1\endcsname\relax
\typeout{** WARNING: IEEEtran.bst: No hyphenation pattern has been}\typeout{** loaded for the language `#1'. Using the pattern for}\typeout{** the default language instead.}\else
\language=\csname l@#1\endcsname
\fi
#2}}
\providecommand{\BIBdecl}{\relax}
\BIBdecl

\bibitem{joelle2020}
J.~Pineau, P.~Vincent-Lamarre, K.~Sinha, V.~Larivière, A.~Beygelzimer,
  F.~d’Alché Buc, E.~Fox, and H.~Larochelle, ``Improving reproducibility in
  machine learning research,'' \emph{Journal of Machine Learning Research
  (JMLR)}, vol.~22, no.~1, pp. 7459--7478, 2021.

\bibitem{newman2009data}
P.~Newman and P.~Corke, ``Data papers—peer reviewed publication of high
  quality data sets,'' \emph{International Journal of Robotics Research
  (IJRR)}, vol.~28, no.~5, pp. 587--587, 2009.

\bibitem{mnist}
Y.~LeCun, C.~Cortes, and C.~J. Burges, ``The {MNIST} dataset of handwritten
  digits,'' available online at \url{http://yann.lecun.com/exdb/mnist} (last
  accessed: 2022-11-13).

\bibitem{imagenet}
J.~Deng, W.~Dong, R.~Socher, L.-J. Li, K.~Li, and L.~Fei-Fei, ``{ImageNet}: A
  large-scale hierarchical image database,'' in \emph{IEEE Conference on
  Computer Vision and Pattern Recognition (CVPR)}, Miami, FL, 2009, pp.
  248--255.

\bibitem{Krizhevsky2009LearningML}
A.~Krizhevsky, ``Learning multiple layers of features from tiny images,'' Tech.
  Rep., 2009, available online at
  \url{https://www.cs.toronto.edu/~kriz/learning-features-2009-TR.pdf} (last
  accessed: 2022-11-13).

\bibitem{paperswithcode}
R.~Stojnic, R.~Taylor, M.~Kardas, E.~Saravia, G.~Cucurull, and T.~Scialom,
  ``{Papers with Code},'' available online at \url{https://paperswithcode.com/}
  (last accessed: 2022-11-13).

\bibitem{quigley2009ros}
M.~Quigley, K.~Conley, B.~Gerkey, J.~Faust, T.~Foote, J.~Leibs, R.~Wheeler, and
  A.~Y. Ng, ``{ROS}: an open-source robot operating system,'' in \emph{IEEE
  International Conference on Robotics and Automation (ICRA) Workshop on Open
  Source Software}, Kobe, Japan, 2009, pp. 1--5.

\bibitem{rtb}
P.~Corke and J.~Haviland, ``Not your grandmother’s toolbox--the robotics
  toolbox reinvented for python,'' in \emph{IEEE International Conference on
  Robotics and Automation (ICRA)}, Xi'an, China, 2021, pp. 11\,357--11\,363.

\bibitem{wang2019benchmarking}
T.~Wang, X.~Bao, I.~Clavera, J.~Hoang, Y.~Wen, E.~Langlois, S.~Zhang, G.~Zhang,
  P.~Abbeel, and J.~Ba, ``Benchmarking model-based reinforcement learning,''
  \emph{arXiv preprint arXiv:1907.02057}, 2019.

\bibitem{how2018control}
J.~P. How, ``Control systems reproducibility challenge [from the editor],''
  \emph{IEEE Control Systems Magazine}, vol.~38, no.~4, pp. 3--4, 2018.

\bibitem{how2015}
------, ``Benchmarks [from the editor],'' \emph{IEEE Control Systems Magazine},
  vol.~35, no.~1, pp. 6--7, 2015.

\bibitem{dodgedrone}
``{ICRA 2022 DodgeDrone Challenge: Vision-based Agile Drone Flight},''
  available online at \url{https://uzh-rpg.github.io/icra2022-dodgedrone} (last
  accessed: 2022-11-13).

\bibitem{iros-comp}
``{IROS 2022 Safe Robot Learning Competition},'' available online at
  \url{https://www.dynsyslab.org/iros-2022-safe-robot-learning-competition}
  (last accessed: 2022-11-13).

\bibitem{the_turing_way2022}
{The Turing Way Community}, ``{The Turing Way: A handbook for reproducible,
  ethical and collaborative research},'' 2022, available online at
  \url{https://doi.org/10.5281/zenodo.7470333} (last accessed: 2022-11-13).

\bibitem{Hirsch2005}
J.~E. Hirsch, ``An index to quantify an individual's scientific research
  output,'' \emph{Proceedings of the National Academy of Sciences}, vol. 102,
  no.~46, pp. 16\,569--16\,572, 2005.

\bibitem{googlescholarh5index}
``Google scholar metrics,'' available online at
  \url{https://scholar.google.com/intl/en/scholar/metrics.html#metrics} (last
  assessed: 2022-12-29).

\bibitem{googlescholarML}
``Top publications: Artificial intelligence,'' available online at
  \url{https://scholar.google.com/citations?view_op=top_venues\&hl=en\&vq=eng_artificialintelligence}
  (last accessed: 2022-12-29).

\bibitem{googlescholarRobotics}
``Top publications: Robotics,'' available online at
  \url{https://scholar.google.com/citations?view_op=top_venues\&hl=en\&vq=eng_robotics}
  (last accessed: 2022-12-29).

\bibitem{googlescholarControl}
``Top publications: Automation and control theory,'' available online at
  \url{https://scholar.google.com/citations?view_op=top_venues&hl=en\&vq=eng_automationcontroltheory}
  (last accessed: 2022-12-29).

\bibitem{Ravenscroft2017}
J.~Ravenscroft, M.~Liakata, A.~Clare, and D.~Duma, ``Measuring scientific
  impact beyond academia: An assessment of existing impact metrics and proposed
  improvements,'' \emph{PLOS ONE}, vol.~12, no.~3, pp. 1--21, 2017.

\bibitem{neurips-num-submissions}
X.~Li, ``Acceptance rates for the \st{major} top-tier {AI}-related
  conferences,'' available online at
  \url{https://github.com/lixin4ever/Conference-Acceptance-Rate} (last
  accessed: 2023-08-11).

\bibitem{icra-num-submissions2}
M.~Q.-H. Meng, Y.~Sun, and J.~Wang, ``A successful hybrid {ICRA} 2021 [society
  news],'' \emph{IEEE Robotics and Automation Magazine}, vol.~28, no.~3, pp.
  182--186, 2021.

\bibitem{cdc-2021-final-program}
``{IEEE Conference on Decision and Control (CDC) 2021}: Final program,''
  available online at
  \url{https://ieeexplore.ieee.org/stamp/stamp.jsp?tp=&arnumber=9683189} (last
  accessed: 2022-12-29).

\bibitem{osc-repo}
A.~Tao, L.~Brunke, S.~Zhou, and J.~Panerati, ``\texttt{code-release}
  repository,'' available online at
  \url{https://github.com/utiasDSL/code-release} (last accessed 2023-08-15).

\bibitem{icraCfP2020}
``{IEEE International Conference on Robotics and Automation (ICRA) 2020}: Call
  for papers,'' available online at
  \url{https://ewh.ieee.org/soc/ras/conf/fullysponsored/icra/ICRA2020/www.icra2020.org/call-for-papers.html}
  (last accessed: 2022-12-29).

\bibitem{cdcCfP2021}
``{IEEE Conference on Decision and Control (CDC) 2021}: Call for papers,''
  available online at \url{https://2021.ieeecdc.org/call-for-papers} (last
  accessed: 2022-12-29).

\bibitem{Borges2016}
H.~Borges, A.~Hora, and M.~T. Valente, ``Understanding the factors that impact
  the popularity of {GitHub} repositories,'' in \emph{IEEE International
  Conference on Software Maintenance and Evolution (ICSME)}, Raleigh, NC, 2016,
  pp. 334--344.

\bibitem{Borges2018}
H.~Borges and M.~{Tulio Valente}, ``What’s in a {GitHub} star? understanding
  repository starring practices in a social coding platform,'' \emph{Journal of
  Systems and Software}, vol. 146, pp. 112--129, 2018.

\bibitem{Rousseeuw1999}
P.~J. Rousseeuw and K.~van Driessen, ``A fast algorithm for the minimum
  covariance determinant estimator,'' \emph{Technometrics}, vol.~41, no.~3, pp.
  212--223, 1999.

\bibitem{neuripsCfP2021}
``{Conference on Neural Information Processing Systems (NeurIPS) 2021}: Call
  for papers,'' available online at
  \url{https://neurips.cc/Conferences/2021/CallForPapers} (last accessed:
  2022-12-29).

\bibitem{stanley}
S.~Thrun, M.~Montemerlo, H.~Dahlkamp, D.~Stavens, A.~Aron, J.~Diebel, P.~Fong,
  J.~Gale, M.~Halpenny, G.~Hoffmann, K.~Lau, C.~Oakley, M.~Palatucci, V.~Pratt,
  P.~Stang, S.~Strohband, C.~Dupont, L.-E. Jendrossek, C.~Koelen, C.~Markey,
  C.~Rummel, J.~van Niekerk, E.~Jensen, P.~Alessandrini, G.~Bradski, B.~Davies,
  S.~Ettinger, A.~Kaehler, A.~Nefian, and P.~Mahoney, ``Stanley: The robot that
  won the darpa grand challenge,'' \emph{Journal of Field Robotics}, vol.~23,
  no.~9, pp. 661--692, 2006.

\bibitem{nimbro}
M.~Beul, M.~Nieuwenhuisen, J.~Quenzel, R.~A. Rosu, J.~Horn, D.~Pavlichenko,
  S.~Houben, and S.~Behnke, ``Team {NimbRo} at {MBZIRC} 2017: Fast landing on a
  moving target and treasure hunting with a team of micro aerial vehicles,''
  \emph{Journal of Field Robotics}, vol.~36, no.~1, pp. 204--229, 2019.

\bibitem{cerberus}
M.~Tranzatto, T.~Miki, M.~Dharmadhikari, L.~Bernreiter, M.~Kulkarni,
  F.~Mascarich, O.~Andersson, S.~Khattak, M.~Hutter, R.~Siegwart, and
  K.~Alexis, ``{CERBERUS} in the {DARPA} subterranean challenge,''
  \emph{Science Robotics}, vol.~7, no.~66, p. eabp9742, 2022.

\bibitem{zeus}
K.~Burnett, J.~Qian, X.~Du, L.~Liu, D.~J. Yoon, T.~Shen, S.~Sun, S.~Samavi,
  M.~J. Sorocky, M.~Bianchi, K.~Zhang, A.~Arkhangorodsky, Q.~Sykora, S.~Lu,
  Y.~Huang, A.~P. Schoellig, and T.~D. Barfoot, ``Zeus: A system description of
  the two-time winner of the collegiate {SAE} autodrive competition,''
  \emph{Journal of Field Robotics}, vol.~38, no.~1, pp. 139--166, 2021.

\end{thebibliography}
\end{document}